\documentclass[secnumarabic,amsmath,amssymb,showpacs,nobibnotes,aps, prl,12pt]{revtex4-1}

\usepackage{amssymb}

\usepackage{graphicx}
\usepackage{dcolumn}
\usepackage{bm}
\usepackage{color}

\begin{document}
\title{Anomalous pinch of turbulent plasmas driven by the magnetic-drift-induced Lorentz force through the Stokes-Einstein relation}

\author{Shaojie Wang}
\email{wangsj@ustc.edu.cn}
\affiliation{Department of Modern Physics, University of Science and Technology of China, Hefei, 230026, China}
\date{\today}

\begin{abstract}
It is found that the Lorentz force generated by the magnetic drift drives a generic plasma pinch fluxes of particle, energy and momentum through the Stokes-Einstein relation. The proposed theoretical model applies for both electrons and ions, trapped particles and passing particles.
\end{abstract}

\pacs{52.25.Dg, 52.25.Fi, 52.20.Dq, 52.65.-y}

\maketitle

Anomalous pinch fluxes of particle, energy and parallel momentum are crucially important in understanding the turbulent transport in the magnetic fusion plasma research \cite{AngioniNF12}. Since the neoclassical Ware pinch velocity \cite{WarePRL70} is insufficient to interpret the related experimental observations on particle pinch, anomalous trapped particle pinch due to the inhomogeneity of magnetic field strength is proposed by using the fluid model \cite{WeilandNF89} and by invoking the turbulence equipartition (TEP) theory based on the invariant measure \cite{IsichenkoPRL95}; this anomalous trapped particle pinch has been confirmed by numerical simulations \cite{GarbetPRL03}. However, since the TEP theory of trapped electron pinch depends on the longitudinal invariant, it is not clear whether it applies to the trapped ions and passing particles, and the ambipolarity issue is unresolved \cite{BakerPoP98}. Recently, the parallel momentum pinch is extensively investigated in the magnetic fusion community to understand the intrinsic toroidal rotation \cite{RiceNF07}. A parallel momentum pinch due to the Coriolis force in a toroidally-rotating frame was proposed by using the fluid model to analyze the quasilinear transport induced by the ion temperature gradient mode \cite{PeetersPRL07}. However, the effects of density gradient and Coriolis force have not been resolved in the momentum pinch theory \cite{HahmPoP08,PeetersPoP09,HahmPoP09}.

Here we propose a new theory to unify the particle, energy and parallel momentum pinch fluxes based on different physical mechanisms. It is found that the Lorentz force generated by the magnetic drift drives a generic pinches through the celebrated Stokes-Einstein relation \cite{Einstein1905a}. The proposed model applies for both electrons and ions, both trapped particles and passing particles; it clearly resolves the above important controversial issues in anomalous pinch.

Consider an axisymmetric tokamak configuration. The equilibrium magnetic filed is given by
\begin{equation}
\bm B=I\left(r\right) \nabla \phi + \nabla \phi\times \nabla r \psi'\left(r\right),
\end{equation}
with $r$ the effective minor radius, $\theta$ the poloidal angle and $\phi$ the toroidal angle. $\psi\left(r\right)$ is the poloidal magnetic flux labeled by $r$, and the prime denotes derivative with respect to $r$.
In the equilibrium field, the guiding-center velocity is given by

\begin{equation}
\bm V_{gc}=v_{\parallel}\bm b+ \bm V_{\bm{E}\times \bm{B}}+\bm{V}_{\nabla B}+\bm {V}_{cur},
\end{equation}
with $v_{\parallel}$ the parallel velocity, $\bm b= \bm{B}/B$; the well-known $\nabla B$ drift and curvature drift are given by
\begin{equation}
\bm{V}_{\nabla B}+\bm {V}_{cur}=\frac{-\mu \nabla B -mv_{\parallel}^2\bm{\kappa}}{eB^2}\times \bm{B}, \label{eq:Vd}
\end{equation}
with $\mu$ the guiding-center magnetic moment. $\bm{\kappa}$ is the curvature of the magnetic field line. $\bm V_{\bm{E}\times \bm{B}}=\bm{E}\times \bm{B}/B^2$, with $\bm{E}=-\nabla \Phi\left(r\right)$ the equilibrium radial electric field, which satisfies $\bm{B}\cdot\bm{E}=0$.

Since the guiding-center velocity is the particle velocity averaged over the gyro-period, the force acted on the particle averaged on the time scale longer than the gyro-period is given by
\begin{equation}
\bm{F}_{L}=e\left(\bm{E}+\bm{V}_{gc}\times\bm{B}\right)=m v_{\parallel}^2\bm{\kappa}+\mu \nabla B. \label{eq:LF-p}
\end{equation}
We point out that the radial electric field force is canceled by the $e\bm{V}_{\bm{E}\times \bm {B}}\times \bm{B}$ force, the mean radial electric field does not play a role here. This charge independent Lorentz force drives the pinch fluxes through the Stokes-Einstein relation \cite{Einstein1905a}, as will be shown in the following. The readers, who are familiar with the Stokes-Einstein relation and not interested in the detailed derivation, may skip to Eq. (\ref{eq:Fa}).

Consider a radial random walk due to the low-frequency nonlinear electrostatic turbulence. Here, the low-frequency means that $\left(\omega_{wave},1/\tau_c\right)\ll\Omega$, with $\omega_{wave}$ the characteristic frequency of the wave, $\tau_c$ the correlation time of the turbulence, and $\Omega$ the gyro-frequency of the charged particle. The work done by the Lorentz force along the random radial displacement $\delta r$ taking place on a time scale longer than the gyro-period changes the particle kinetic energy by
\begin{equation}
\delta \mathcal{K}=\bm{F}_{L}\cdot \nabla r \delta r=\left(m v_{\parallel}^2\bm{\kappa}+\mu \nabla B\right)\cdot \nabla r \delta r, \label{eq:dK}
\end{equation}
which is again charge independent. This equation should not be confused with $e\left(\bm{V}_{gc}\times\bm{B}\right)\cdot\bm{V}_{gc}=0$.

The particle kinetic energy is given by
\begin{equation}
\mathcal {K}=\left(1/2\right)m v_{\parallel}^2+\mu B.  \label{eq:K}
\end{equation}

From Eq. (\ref{eq:dK}) and Eq. (\ref{eq:K}), one finds
\begin{equation}
\delta v_{\parallel}= v_{\parallel}\bm{\kappa}\cdot \nabla r \delta r, \label{eq:dv}
\end{equation}
which follows from
\begin{equation}
\delta \mathcal{K}=m v_{\parallel} \delta v_{\parallel} + \mu \partial_r B \delta r. \label{eq:dK1}
\end{equation}

We point out that Eq. (\ref{eq:dv}) can also be derived in a guiding-center picture. Note that the guiding-center kinetic energy is
\begin{equation}
\mathcal{K}_{gc}=\left(1/2\right)mv_{\parallel}^2,
\end{equation}
and $\mu B$ is the well-known magnetic potential energy of a guiding-center due to its constant magnetic moment $\mu$.
The force acted on a guiding-center is therefore given by
\begin{equation}
\bm{F}_{gc}=e\left(\bm{E}+\bm{V}_{gc}\times\bm {B} \right)-\mu\nabla B=m v_{\parallel}^2\bm{\kappa},
\end{equation}
which is simply the centripetal force that is necessary to maintain the parallel motion along the field line; this indicates that the centripetal force needed to maintain the parallel streaming is provided by the Lorentz force generated by the curvature drift.
Using
\begin{equation}
\delta \mathcal {K}_{gc}=\bm{F}_{gc}\cdot \nabla r \delta r,
\end{equation}
one finds again Eq. (\ref{eq:dv}), which clearly predicts the cross-correlation between $\delta r$ and $\delta v_{\parallel}$.

Let $f \left(\bm z, t\right)$ denote the ensemble averaged particle distribution in the phase space $\bm z=\left(r,\theta,v_{\parallel},\mu\right)$.
The phase space transport equation for a nonlinear turbulent plasma is given by a Fokker-Planck equation \cite{BakerPoP98,ChenJGR99,KominisPRL10,WangPoP12,WangPoP16},
\begin{equation}
\partial_t  f -\frac{1}{\mathcal J}\partial_i\left(\mathcal J d^{ij} \partial_j f \right)=0,\label{eq:FP}
\end{equation}
with $\mathcal J$ the Jacobian of the phase-space, and the phase space diffusivity is given by
\begin{equation}
d^{ij}\equiv\left\langle \delta z^i \delta z^j \right\rangle _ {en}/2\delta t, \label{eq:phase-D}
\end{equation}
where $\langle \cdot \rangle_{en}$ denotes the ensemble average. Note that $\delta z^i$ denotes the departure of the particle from the unperturbed orbit within a time interval $\delta t$ which is longer than $\tau_c$, and in a tokamak the unperturbed motion does not contribute to transport by itself.
In Eq. (\ref{eq:FP}), $i, j$ stands for $r$ and $v_{\parallel}$; note that $\delta \mu =0$ due to the conservation of magnetic moment, and $\delta \theta$ is ignored since we shall concentrate on the radial transport.

Using Eq. (\ref{eq:dv}) one finds
\begin{equation}
d^{rv_{\parallel}}=d^{rr}\nabla r \cdot \bm{\kappa} v_{\parallel}, \label{eq:drv}
\end{equation}
\begin{equation}
d^{v_{\parallel}v_{\parallel}}=v_{\parallel}\bm{\kappa}\cdot \nabla r  d^{rr}\nabla r \cdot \bm{\kappa}v_{\parallel}. \label{eq:dvv}
\end{equation}

To proceed, we assume that the ensemble-averaged particle distribution is a local shift Maxwellian \cite{HahmPoP08},
\begin{equation}
 f=\frac{n}{\left(2\pi T/m\right)^{\frac{3}{2}}}\exp\left(-\frac{\frac{1}{2}m\left(v_{\parallel}-U\right)^2+\mu B}{T}\right),\label{eq:dist}
\end{equation}
with $n$ and $T$ the particle density and temperature, respectively. $U$ is the fluid parallel velocity; $p=nT$ is the pressure. $m$ and $e$ are the particle mass and electrical charge, respectively. Note that the particle species index is dropped in the above formulation; clearly, our discussion applies for both ions and electrons.

Define a set of forces
\begin{subequations}
\begin{eqnarray}
A_{1}&\equiv&-\partial_r \ln p,\\
A_{2}&\equiv&-\partial_r \ln T,\\
A_{3}&\equiv&-\partial_r \ln U,\\
A_{4}&\equiv&\partial_r \ln B,\\
A_{5}&\equiv& \nabla r \cdot\bm{\kappa}.
\end{eqnarray}
\end{subequations}

Introduce a set of weighting functions
\begin{subequations}
\begin{eqnarray}
h_1&\equiv&1,\\
h_{2}&\equiv&\frac{\frac{1}{2}m\left(v_{\parallel}-U\right)^2+\mu B}{T}-\frac{5}{2},\\
h_{3}&\equiv&mU\left(v_{\parallel}-U\right)/T,\\
h_{4}&\equiv&\mu B/T,\\
h_{5}&\equiv&m v_{\parallel}\left(v_{\parallel}-U\right)/T.
\end{eqnarray}
\end{subequations}

Using the above definitions, one finds
\begin{equation}
  \partial_t  f +\frac{1}{\mathcal J}\partial_r\left[\mathcal J \left(d^{rr} h_{\alpha}A_{\alpha} \right) f  \right]+\frac{1}{\mathcal J}\partial_{v_{\parallel}}\left[\mathcal J \left(v_{\parallel} A_5 d^{rr} h_{\alpha}A_{\alpha} \right) f  \right]=0,\label{eq:FP1}
\end{equation}
where and in the following $\alpha, \beta=1,..., 5$ is understood.

Note that $A_1$, $A_2$, $A_3$ are the usual thermodynamic forces; $A_4$ stands for the Lorentz force due to the $\nabla B$ drift, and $A_5$ stands for the Lorentz force due to the curvature drift. In a large-aspect-ratio tokamak, we have $R=R_0+r\cos\theta$, with $r\ll R_0$, here $R_0$ is the major radius of the torus. Using the large-aspect-ratio approximation, one finds $B\propto 1/R$, $\bm{\kappa}\approx -\bm{R}/R^2$, and
\begin{equation}
A_{4}\approx -\frac{1}{R}\cos\theta \approx A_{5}.
\end{equation}

Using Eq. (\ref{eq:FP}), one finds the local production rate of the entropy density,
\begin{equation}
\sigma=\int d^3\bm v  f \partial_i \ln f d^{ij}\partial_j \ln f. \label{eq:entropy}
\end{equation}

Using Eq. (\ref{eq:FP1}), one writes the entropy density production rate as a positive-definite quadratic form \cite{WangPoP16}
\begin{equation}
\sigma=A_{\alpha}J_{\alpha },
\end{equation}
with the canonical conjugate fluxes given by
\begin{equation}
J_{\alpha}=L_{\alpha \beta}A_{\beta},\label{eq:flux}
\end{equation}
and the transport matrix is given by
\begin{equation}
L_{\alpha \beta}=\int d^3 \bm{v} f d^{rr} h_{\alpha}h_{\beta}\equiv n\left\langle h_{\alpha}d^{rr}h_{\beta}\right\rangle, \label{eq:tmat}
\end{equation}
which satisfies the Onsager symmetry relation \cite{OnsagerPR31a,OnsagerPR31b}.

To proceed, one needs to find the magnetic-flux-surface averaged radial transport equations, which are obtained by integrating Eq. (\ref{eq:FP1}) multiplied by $\left[1, \frac{1}{2}m\left(v_{\parallel}-U\right)^2+\mu B, mv_{\parallel}\right]$ over the velocity space and averaging the results over the magnetic flux surface.

The particle transport equation is
\begin{equation}
\partial_t  n+\frac{1}{V'\left(r\right)}\partial_r \left[V'\left(r\right)\Gamma^r\right]=0,\label{eq:part-t}
\end{equation}
with $V\left(r\right)$ the volume enclosed in the flux-surface labeled by $r$, and the radial particle flux $\Gamma^r$ given by
\begin{equation}
\Gamma^r=J_{1}.
\end{equation}

The energy transport equation is
\begin{equation}
\partial_t \left( \frac{3}{2}p\right)+\frac{1}{V'} \partial_r \left[V'\left(q^r+\frac{5}{2}\Gamma^r T\right)\right]=Q,\label{eq:ener-t}
\end{equation}
with $q^r$, the radial heat flux given by
\begin{equation}
q^r/T=J_{2},
\end{equation}
and $Q$, the turbulence heating rate given by
\begin{equation}
Q/T=J_{3}A_{3}+J_{4}A_{4}+J_{5}A_{5}. \label{eq:t-h}
\end{equation}

The parallel momentum transport equation is
\begin{equation}
\partial_t \left( n m  U\right)+\frac{1}{V'}\partial_r \left[V'\left(\Pi_{\parallel}^r+ \Gamma^r m U\right)\right]=F_{\parallel},\label{eq:momentum-t}
\end{equation}
with $\Pi_{\parallel}^r$, the radial component of the parallel viscosity given by
\begin{equation}
\Pi_{\parallel}^rU/T=J_{3},\label{eq:visc}
\end{equation}
and $F_{\parallel}$, the turbulence parallel acceleration rate given by
\begin{equation}
F_{\parallel}U/T=\left(J_{3}+J_{1}mU^2/T\right)A_{5}. \label{eq:t-a}
\end{equation}

Note that in the above flux-surface averaged transport equations, $A_4=-f_{balloon}/R=A_5$ should be understood, with the ballooning factor $f_{balloon}\approx 1$ for the strong ballooning turbulence \cite{HahmPoP08}.

Note that $L_{11}=nD$, $L_{22}=n\chi$, and $L_{33}=n\chi_{\phi}mU^2/T$,
with $D$, $\chi$, and $\chi_{\phi}$ the usual particle diffusivity, heat diffusivity, and toroidal viscosity, respectively.
In the following, we shall use the constant $d^{rr}$ approximation \cite{DupreePF67} for mathematical simplicity.  This approximation gives $D=d^{rr}$, $\chi=\frac{5}{2}d^{rr}$, and $\chi_{\phi}=d^{rr}$.

To evaluate the particle pinch due to the $\nabla B$ drift ($L_{14}A_4$) and the curvature drift ($L_{15}A_5$), we write down
\begin{equation}
L_{14}=n\left\langle d^{rr}h_1h_4\right\rangle= L_{11}, \label{eq:gradB-p}
\end{equation}
\begin{equation}
L_{15}=n\left\langle d^{rr}h_1h_5\right\rangle= L_{11}. \label{eq:curv-p}
\end{equation}
Note that the previous trapped electron pinch \cite{IsichenkoPRL95,BakerPoP98} is included in Eq. (\ref{eq:gradB-p}) (the effect of $\nabla B$ drift). We point out that Eq. (\ref{eq:curv-p}) indicates that curvature drift induces a particle pinch. Eqs. (\ref{eq:gradB-p}, \ref{eq:curv-p}) are also applicable for passing particles. Therefore, the present theory clearly predicts a passing particle pinch, which is absent in previous fluid theory \cite{WeilandNF89} and kinetic theory \cite{IsichenkoPRL95,BakerPoP98}. It should be noted that in a more recent fluid theory and simulation work \cite{AngioniPRL06}, particle pinch of impurity ions was identified, which is independent of the trapping fraction, as is different from the earlier fluid theory \cite{WeilandNF89}; however, it is not clear in Ref. \onlinecite{AngioniPRL06} whether the result can be extended to the primary ions, partly because the role of trapped particles and passing particles is not clear in a fluid theory. Since the particle pinch found here is driven by the Lorentz force due to the magnetic drift, it applies for both ions and electrons; there is no need of an ambipolar radial electric field discussed in Ref. \onlinecite{BakerPoP98} where only the trapped electron pinch was identified by invoking the argument of invariant measure.

The constant $d^{rr}$ approximation gives $L_{12}=-L_{11}$, $L_{13}=0=L_{23}$. The particle flux can be written in a more familiar form
\begin{equation}
\Gamma^r=-D\partial_r n +nV_{p,pinch},
\end{equation}
with the particle pinch velocity given by
\begin{equation}
V_{p,pinch}=-2\frac{D}{R}. \label{eq:p-p}
\end{equation}

To evaluate the energy pinch due to the $\nabla B$ drift and the curvature drift, we write down
\begin{equation}
L_{24}=n\left\langle d^{rr}h_2h_4\right\rangle= 0, \label{eq:gradB-e}
\end{equation}
\begin{equation}
L_{25}=n\left\langle d^{rr}h_2h_5\right\rangle =0. \label{eq:curv-e}
\end{equation}

The total energy flux can be written as
\begin{equation}
q^r+\frac{5}{2}\Gamma^r T=-\frac{3}{5}n\chi\partial_r T-\frac{3}{2}pD\partial_r \ln n+\frac{3}{2}p V_{\mathcal{E},pinch},
\end{equation}
with the energy pinch velocity given by
\begin{equation}
V_{\mathcal{E},pinch}=-\frac{4}{3}\frac{\chi}{R}.\label{eq:e-p}
\end{equation}
Note that the energy pinch thus derived is contributed by the particle pinch.

To evaluate the parallel momentum pinch due to the $\nabla B$ drift and the curvature drift, we write down
\begin{equation}
L_{34}=n\left\langle d^{rr}h_3h_4\right\rangle = 0, \label{eq:gradB-m}
\end{equation}
\begin{equation}
L_{35}=n\left\langle d^{rr}h_3h_5\right\rangle = L_{33}. \label{eq:curv-m}
\end{equation}
To complete the computation of the transport matrix, we list $L_{44}=2L_{11}$, $L_{45}=L_{11}$, $L_{55}=\left(3+mU^2/T\right)L_{11}$.

The total parallel momentum flux can be written as
\begin{equation}
\Pi_{\parallel}^r+ \Gamma^r m U=-nm\chi_{\phi}\partial_r U -nmUD\partial_r \ln n +nmUV_{\phi,pinch},
\end{equation}
with the parallel momentum pinch velocity
\begin{equation}
V_{\phi,pinch}=-3\frac{\chi_{\phi}}{R}, \label{eq:m-p}
\end{equation}
which agrees with the previous result \cite{HahmPoP08}. The present theory indicates that two third of this momentum pinch comes from the particle pinch due to both the $\nabla B$ drift ($L_{14}A_4$ term) and the curvature drift ($L_{15}A_5$ term), and one third of the momentum pinch comes from the $L_{35}A_5$ term (effect of curvature drift) in the parallel viscosity; in another word, two third of this momentum pinch is from the curvature drift effect ($L_{15}A_{5}$ term and $L_{35}A_{5}$ term), and one third of this pinch is from the $\nabla B$ drift effect ($L_{14}A_{4}$ term). Note that the density gradient effect on the convection of momentum comes from the particle flux, which is always outward in the usual central peaking density case; on this point, we agree with Refs. \onlinecite{HahmPoP08,HahmPoP09}, disagree with Refs. \onlinecite{PeetersPRL07,PeetersPoP09}. Note also that a parallel acceleration/deceleration term due to the curvature drift [Eq. (\ref{eq:t-a}) or $F_{\parallel}=\left(\Pi_{\parallel}^r+\Gamma^rmU\right)A_{5}$] is missing in the previous momentum pinch theories \cite{PeetersPRL07,HahmPoP08}.

To further clarify the physical mechanism of the anomalous pinch, we introduce
\begin{equation}
F_{a}=-f_{balloon}\frac{2T}{R},\label{eq:Fa}
\end{equation}
which is the averaged radial Lorentz force generated by the $\nabla B$ drift and the curvature drift according to Eq. (\ref{eq:LF-p}). Here we have restored the ballooning factor.
The pinch velocities of particle, energy and parallel momentum are written as
\begin{equation}
V_{p,pinch}=F_{a}\frac{D}{T}, \label{eq:p-p1}
\end{equation}
\begin{equation}
V_{\mathcal{E},pinch}=\frac{2}{3}F_{a}\frac{\chi}{T}, \label{eq:e-p1}
\end{equation}
\begin{equation}
V_{\phi,pinch}=\frac{3}{2}F_{a}\frac{\chi_{\phi}}{T}, \label{eq:m-p1}
\end{equation}
respectively.
Eq. (\ref{eq:p-p1}) agrees with the celebrated Stokes-Einstein relation \cite{Einstein1905a,PathriaBook12}, which indicates that the particle pinch is driven by the Lorentz force generated by the magnetic drift. Eqs. (\ref{eq:p-p1}, \ref{eq:e-p1}, \ref{eq:m-p1}) shall be referred to as the generalized perpendicular Stokes-Einstein relation, which describes the anomalous pinch velocities of particle, energy and momentum driven by the Lorentz force induced by the magnetic drift in a generic form. Note that the parallel Stokes-Einstein relation in a turbulent plasma has been previously discussed \cite{WangPoP16}.

In conclusion, we have proposed a new theoretical model based on the generalized Stokes-Einstein relation to unify the physical pictures of the anomalous pinch of particle, energy and parallel momentum, which is accomplished by identifying for the first time the averaged radial Lorentz force generated by the $\nabla B$ drift and the curvature drift. The new simple mechanism of anomalous pinch covers the previous trapped electron pinch due to the conservation of longitudinal invariant and the previous parallel momentum pinch due to the Coriolis force in a rotating frame. The proposed theory has resolved several important issues in anomalous pinch.

(1) It predicts the effect of passing particle pinch, which is absent in previous TEP theory \cite{IsichenkoPRL95,BakerPoP98}.

(2) It predicts the particle pinch for both electrons and ions, while the previous theory applies only for trapped electrons and an ambipolar radial electric field is needed to adjust the ion particle flux according to Baker-Rosenbluth's viewpoint \cite{BakerPoP98}. Eq. (\ref{eq:LF-p}) indicates that the perpendicular mean electric field does not drive a particle pinch through the Stokes-Einstein relation, due to the cancelation between $e\bm {E}$ and $e\bm{V}_{\bm{E}\times\bm{B}}\times\bm{B}$, therefore, the anomalous particle pinch must be auto-ambipolar.

(3) It predicts that the inward-directed density gradient always drives an outward momentum flux, which agrees with Ref. \onlinecite{HahmPoP09}, disagrees with Ref. \onlinecite{PeetersPoP09}.

(4) The physical picture of parallel momentum pinch described in this paper is related to the Stokes-Einstein flux driven by the Lorentz force generated by the magnetic drift, which is different from the previous understanding by invoking the Coriolis force \cite{PeetersPRL07,HahmPoP08,PeetersPoP09,HahmPoP09}. Note the fact that the Coriolis force is in a rotating frame, it does not exist in the laboratory frame, however, the radial transport fluxes should not be changed by simply switching from the laboratory frame \cite{HahmPoP08,HahmPoP09} to a toroidally-rotating frame \cite{PeetersPRL07,PeetersPoP09}. Note that the Lorentz force proposed here is independent of the choice of reference frame.

It is of interest to note that the Stokes-Einstein fluxes discussed in this paper is from the cross-correlation in the phase-space ($\left\langle\delta r \delta v_{\parallel}\right\rangle_{en}$, $\left\langle\delta r \delta \mathcal{K}\right\rangle_{en}$, etc.), as was pointed out previously \cite{WangPoP16}. The method developed here may also be extended to the magnetic turbulence in space plasmas \cite{ChenJGR99} and to the general statistical physics \cite{PathriaBook12}.

\begin{acknowledgments}
This work was supported by the National Natural Science Foundation of China under Grant No. 11375196 and the National ITER program of China under Contract No. 2014GB113000.
\end{acknowledgments}

\nocite{*}

%

\end{document}